\documentclass[fleqn,usenatbib]{mnras}

\usepackage{newtxtext,newtxmath}

\usepackage[T1]{fontenc}

\DeclareRobustCommand{\VAN}[3]{#2}
\let\VANthebibliography\thebibliography
\def\thebibliography{\DeclareRobustCommand{\VAN}[3]{##3}\VANthebibliography}

\usepackage{graphicx}	
\usepackage{amsmath}	
\usepackage{tabularx}

\title[DM estimation with ML]{Machine-learning approaches to dispersion measure estimation for fast radio bursts}

\author[H. Rajabi et al.]{
Hosein Rajabi,$^{1}$
Zhejian Liu,$^{2}$
Fereshteh Rajabi,$^{2}$
and Martin Houde$^{3}$\thanks{E-mail: mhoude2@uwo.ca}
\\
$^{1}$Department of Computer Science, The University of Western Ontario, 1151 Richmond Street, London, Ontario N6A 3K7, Canada\\
$^2$Department of Physics and Astronomy, McMaster University, 1280 Main Street West, Hamilton, Ontario, L8S 4L8, Canada\\
$^3$Department of Physics and Astronomy, The University of Western Ontario, 1151 Richmond Street, London, Ontario N6A 3K7, Canada
}

\date{Accepted XXX. Received YYY; in original form ZZZ}

\pubyear{\the\year{}}

\begin{document}
\label{firstpage}
\pagerange{\pageref{firstpage}--\pageref{lastpage}}
\maketitle

\begin{abstract}
Fast radio bursts (FRBs) are bright, mostly millisecond-duration transients of extragalactic origin whose emission mechanisms remain unknown. As FRB signals propagate through ionized media, they experience frequency-dependent delays quantified by the dispersion measure (DM), a key parameter for inferring source distances and local plasma conditions. Accurate DM estimation is therefore essential for characterizing FRB sources and testing physical models, yet current dedispersion methods can be computationally intensive and prone to human bias. In this proof-of-concept study, we develop and benchmark three deep-learning architectures, a conventional convolutional neural network (CNN), a fine-tuned ResNet-50, and a hybrid CNN–LSTM model, for automated DM estimation. All models are trained and validated on a large set of synthetic FRB dynamic spectra generated using CHIME/FRB-like specifications. The hybrid CNN–LSTM achieves the highest accuracy and stability while maintaining low computational cost across the investigated DM range. Although trained on simulated data, these models can be fine-tuned on real CHIME/FRB observations and extended to future facilities, providing a pathway toward real-time, data-driven DM estimation in large FRB surveys with further development.
\end{abstract}



\section{Introduction}\label{sec:Introduction}

Fast radio bursts (FRBs) are highly energetic, short-duration radio transients whose physical origin and emission mechanisms remain elusive despite significant theoretical and observational progress. Their spectra provide key diagnostics for probing both the source and its environment. FRBs exhibit several characteristic spectro-temporal features, including the well-known dispersion sweep, where lower frequencies arrive later; a tendency for shorter burst durations at higher observing frequencies; and, in many repeating sources, a systematic downward drift of emission frequency with time across sub-bursts. The observed spectra therefore encode information about both the source and the propagation path from the emission region to the observer.

As the signal propagates, it interacts with intervening media and experiences dispersion, scattering, and scintillation. Disentangling these effects is essential for isolating the intrinsic properties of the source. Accurate dedispersion is particularly critical, as many parameters derived from FRB observations are sensitive to errors in the estimated DM ( \citealt{rajabi2020,chamma2021}). Precise DM estimation not only enables recovery of intrinsic temporal and spectral structures but also provides insights into their host environments.

Once propagation effects are characterized, theoretical models can focus on how intrinsic source properties and dynamics produce the observed spectro-temporal behaviour. The Triggered Relativistic Dynamical Model (TRDM; \citealt{rajabi2020}) is one such framework, quantifying how source dynamics influence spectral structures. TRDM predicted an inverse relationship between sub-burst slope and duration, known as the sub-burst slope law, a prediction subsequently confirmed across multiple repeating FRB sources \citep{rajabi2020,chamma2021,wang2022,jahns2022,chamma2023,Brown2024,2024Chamma}. Such models establish a physical link between source dynamics and observed signatures, offering stringent constraints on theoretical descriptions of FRB emission.

In parallel, considerable effort has focused on developing accurate and efficient dedispersion techniques. In FRB detection pipelines, identifying genuine bursts among false positives (e.g. noise and radio-frequency interference, RFI) involves several processing stages, including real-time dedispersion algorithms that typically maximize the signal-to-noise ratio (S/N) of the dedispersed pulse. 

Two main dedispersion approaches are used in real-time analysis: incoherent and coherent. The incoherent method corrects for dispersion by applying time-delay adjustments to individual frequency channels, with accuracy limited by channel bandwidth. Although less computationally demanding than its coherent counterpart, blind searches over large DM ranges remain computationally intensive. Various implementations, such as brute-force \citep{Barsdell2010MNRAS}, tree dedispersion \citep{Taylor1974}, sub-band methods, and fast discrete dispersion measure transforms (FDMT), offer different trade-offs in accuracy and computational efficiency depending on the underlying hardware architecture (CPU or GPU). Modern optimized pipelines, such as bonsai \citep{CHIME2018}, as well as more recent developments \citep{Men2024}, have demonstrated efficient real-time dedispersion over large DM ranges, including implementations capable of real-time performance on CPU-based systems.

Coherent dedispersion \citep{Hankins1975,Ait2009,VanStraten2011,Bassa2017} operates directly on raw voltage data, reversing the dispersive effects of the interstellar medium (ISM) in the Fourier domain to recover intrinsic pulse shapes with high accuracy. Although particularly effective for high-DM bursts from known repeaters and at very low radio frequencies, it remains prohibitively expensive for blind searches. Semi-coherent approaches, which apply coherent dedispersion at selected DM trials followed by incoherent refinement, can balance sensitivity and computational cost.

While these real-time algorithms enable rapid FRB detection, they are optimized primarily for S/N rather than for fine-scale structural fidelity, and their precision degrades for complex or low-S/N bursts. In contrast, several post-processing algorithms, such as \textit{DM phase} \citep{Seymour2019}, the \textit{DM power} method \citep{Lin2022}, and related structure-maximization techniques \citep{Platts2021,Marthi2022}, focus on recovering the DM that best preserves the burst substructure. These approaches can achieve sub–pc cm$^{-3}$ precision but are computationally intensive and typically applied to bright, high-S/N events in offline analyses.

Despite these advances, dedispersion remains computationally demanding for real-time applications and can lose precision for faint or morphologically complex bursts, as often encountered with instruments such as the Canadian Hydrogen Intensity Mapping Experiment (CHIME/FRB). Moreover, high-precision post-processing methods are impractical for large-scale or automated analyses due to their computational cost and reliance on manual intervention. These limitations motivate the development of data-driven approaches that combine the speed and automation of real-time systems with the accuracy and structural sensitivity of post-processing algorithms.

In this paper, we investigate the use of machine-learning (ML) techniques to improve dedispersion and DM estimation for FRBs. While convolutional neural networks (CNNs; \citealt{connor2018applying}) have previously been proposed as substitutes for traditional dedispersion backends, most ML studies to date have focused on the classification or detection of FRBs rather than on quantitative DM estimation. Here, we extend this exploration by developing and comparing three deep-learning architectures, a baseline CNN, a fine-tuned ResNet-50, and a hybrid CNN–LSTM model, designed to recover the intrinsic DM directly from frequency–time dynamic spectra. Section~\ref{sec:ML_pastwork} reviews previous ML applications to FRBs, Section~\ref{sec:data} describes the generation of synthetic FRB data used for training and validation, and Section~\ref{sec:model_training} outlines the architecture and training procedures for the three models. Our results are presented and discussed in Section~\ref{sec:results}, and the main conclusions and future directions are summarised in Section~\ref{sec:conclusion}. Additional details on the simulated data, cross-validation, and test-set performance are provided in the Appendices at the end.

\section{FRBs and Machine Learning}\label{sec:ML_pastwork}

As previously stated, ML techniques have increasingly been applied to FRBs, though most studies have focused on classification rather than quantitative parameter estimation. Existing applications fall into two main categories: distinguishing genuine FRB signals from RFI or known pulsars \citep{Wagstaff_2016, Akeret2017, Mosiane2017, Czech2018, agarwal2020}, and differentiating between repeating and non-repeating sources \citep{Yang2023, ZhuGe2023, Luo2023, Garcia2024, Sharma2024, Qiang2025, Saliwanchik2022,Sun2025,Sun2026}. In addition, recent approaches based on object-detection frameworks, such as CentreNet \citep{ZHANG2025}, have been proposed to identify burst locations in time-DM space and estimate arrival time and DM, in combination with classification models, further expanding the range of ML techniques applied to FRB analysis. Various ML models have been explored in these contexts. For example, \citet{Wagstaff_2016} used a Random Forest classifier to enhance FRB detection within the VLBA’s V-FASTR project \citep{Wayth_2011}, while \citet{Akeret2017} applied CNNs to discriminate genuine FRBs from RFI and pulsars. \citet{Saliwanchik2022} further used neural networks to classify FRBs as repeating or non-repeating.

A notable example is the work of \citet{agarwal2020}, who demonstrated the use of CNNs and transfer learning to improve FRB detection. They fine-tuned pre-trained image-classification networks such as VGG16 \citep{vgg}, ResNet50 \citep{resnet}, and Xception \citep{xcep} on a combined dataset of simulated FRBs, real RFI events from the Green Bank Observatory, and pulsar data. Their approach incorporated both frequency–time and DM–time representations to capture spectro-temporal and dispersion-related features, introducing a fusion technique that combined information from the two input domains to enhance classification accuracy.

Interest in ML-driven FRB analysis continues to grow, but the development of more sophisticated models has been limited by the scarcity of publicly available, high-quality data. To overcome this, many studies have generated synthetic data to supplement real observations. For instance, \citet{connor2018applying} developed a deep CNN trained on dynamic spectra, pulse profiles, and DM–time data using simulated FRBs injected into actual survey noise. They also explored whether CNNs could eventually replace traditional real-time dedispersion backends, but emphasized that low signal-to-noise per pixel remains a fundamental limitation for such applications with current telescopes.

In this work, we aim to extend ML applications beyond classification by developing models that can directly infer the DM from frequency–time dynamic spectra. Applied immediately after RFI removal, such models could, with further development, automate and refine DM selection, improving the recovery of intrinsic FRB structures and enhancing subsequent spectro-temporal analyses. This study serves as a proof of concept, demonstrating the feasibility of quantitative DM estimation with deep learning and establishing a framework for applying these methods to real observational data in the future.

\section{Data}\label{sec:data}

The models were trained on a synthetic benchmark dataset containing 180,000 samples, randomly divided into six folds for cross-validation. Here, the term “folds” refers to the partitions used in k-fold cross-validation, following standard machine learning terminology. In each iteration, one fold (30,000 samples) was used for validation, while the remaining five were used for training. The validation set was employed to assess model robustness and optimize hyperparameters. An independent test set of 10,000 samples, generated using the same parameter distributions, was used to evaluate performance on unseen data. All datasets follow a uniform DM distribution (see Figure~\ref{fig:dm_histograms}), ensuring balanced coverage across dispersion measures and preventing biases that could affect generalization accuracy.

The synthetic dataset was generated using a publicly available software package hosted on GitHub\footnote{\url{https://github.com/liamconnor/single_pulse_ml}} (GPL-2.0 license), which we extended to more closely reproduce real observations, including implementation of the sub-burst slope law (see below). The modified framework allows simulations with different basic instrumental parameters (e.g., frequency range and bandwidth), but does not explicitly account for site-specific RFI or detailed receiver characteristics.

\subsection{Data Simulation}\label{sec:data-simulation}

To simulate realistic FRB dynamic spectra (frequency–time waterfalls), we generated both the background noise and the FRB signal using parameter distributions consistent with observational data. Telescope characteristics, such as frequency bandwidth, frequency resolution, and sampling time, were incorporated to ensure that the simulated signals captured realistic instrumental effects. In particular, the time and frequency resolutions were not treated as free parameters, but were chosen to be broadly representative of CHIME instrumental characteristics, with a somewhat coarser effective time sampling adopted in the simulations.

For the simulation, we consider a frequency band spanning $\nu_\mathrm{min}=400$~MHz to $\nu_\mathrm{max}=800$~MHz, with a reference frequency $\nu_\mathrm{ref}=600$~MHz, consistent with CHIME/FRB observations. We adopt an effective sampling time of $t_\mathrm{samp}=1.67$~ms, chosen to be broadly representative of the temporal resolution within this observational setup. Any burst intrinsically shorter than this duration will therefore appear broadened to at least $1.67$~ms owing to the finite instrumental sampling time. This temporal resolution also determines the maximum dispersion measure ($\mathrm{DM}_\mathrm{max}$) that can be accommodated within the simulated time window:
\begin{align}
\mathrm{DM}_\mathrm{max} = N_\mathrm{time}\!\left(\frac{t_\mathrm{samp}}{a}\right)\!
\left(\frac{1}{\nu_\mathrm{min}^2}-\frac{1}{\nu_\mathrm{max}^2}\right)^{-1},
\label{eq:DMmax}
\end{align}
where $a = 4.148\,808~\mathrm{GHz^2\,cm^3\,pc^{-1}\,ms}$ \citep{Kulkarni2020}, and $N_\mathrm{time}$ is the number of time bins employed in the simulation. We adopt $N_\mathrm{freq} \times N_\mathrm{time}=512\times 6990$, corresponding to a simulated DM range of $0 \leq \mathrm{DM} \leq 600~\mathrm{pc\,cm^{-3}}$. In practice, we increased the time span by 5\,per\,cent to ensure that the pulse profile is fully covered in each frequency channel, increasing the final number of time bins to 7340.

Observed FRBs exhibit dispersion measures spanning approximately $90~\mathrm{pc\,cm^{-3}}$ \citep{Bhardwaj2021} to $3300~\mathrm{pc\,cm^{-3}}$ \citep{Crawford2022}, with the majority lying below $1000~\mathrm{pc\,cm^{-3}}$ \citep{Spanakis2021}. Since most CHIME/FRB detections fall within this interval \citep{Rafiei2021}, our chosen range represents a statistically significant portion of the observed FRB population while maintaining computational efficiency. This range extends earlier proof-of-concept simulations \citep{Rajabi2024} to a broader
and more CHIME/FRB-relevant DM interval.

The background noise was modelled as Gaussian, representing the stochastic fluctuations typically present in real telescope data. Each element of the frequency–time matrix was drawn from a normal distribution with a mean of zero and a standard deviation of unity. This procedure generated a realistic representation of the thermal noise field over which the synthetic FRB signal was subsequently superimposed. We note, however, that modeling the background as Gaussian noise is an idealization and does not capture RFI, which can affect real observations. In practice, dedicated RFI mitigation is applied prior to analysis in FRB search pipelines, and our approach assumes that such preprocessing has been performed.

To simulate the FRB signal, the temporal envelope of the burst at each frequency channel was represented by a Gaussian profile with unit amplitude,
\begin{align}
g_{\nu}(t) = \exp\!\left[-\frac{(t - t_{0})^2}{2W^2}\right],
\label{eq:g(t)}
\end{align}
where $t_{0}$ and $W$ denote the pulse peak time and the Gaussian standard deviation, respectively.
Both quantities are expressed in units of the receiver’s sampling time to define their locations in the simulated frequency–time matrix.
In practice, the effective width $W$ accounts for the combined effects of intrinsic, instrumental, and dispersion-induced broadening,
\begin{align}
W = \sqrt{t_\mathrm{w}^2 + t_\mathrm{samp}^2 + \Delta t_\mathrm{DM}^2},
\label{eq:W}
\end{align}
where $t_\mathrm{w}$ is the intrinsic width of the FRB pulse and $\Delta t_\mathrm{DM}$ represents temporal smearing due to dispersion across finite frequency channels. In our simulations, the intrinsic width $t_\mathrm{w}$ was drawn from a normal distribution with a mean and standard deviation of $\mu = 0.0015~\mathrm{s}$ and $\sigma = 0.0003~\mathrm{s}$, respectively. Any residual dedispersion errors were neglected, while scattering was incorporated explicitly in a subsequent step.

This characteristic timescale also sets the tolerance for acceptable DM errors in the analysis. For an FRB with an intrinsic duration of $\sim1.5$~ms at 1~GHz \citep{Brown2024}, a DM error of $\Delta\mathrm{DM}\approx1.8~\mathrm{pc\,cm^{-3}}$ produces a residual dispersive smearing comparable to the pulse width. When scaled to the CHIME band centre of 600~MHz, this error budget corresponds to $\Delta\mathrm{DM}\approx1.1~\mathrm{pc\,cm^{-3}}$. As shown in the sub-burst slope analysis of the TRDM framework (\citealt{rajabi2020,Rajabi2024}), DM errors exceeding this level can dominate over the intrinsic drift term, washing out the frequency-time slopes used to constrain source dynamics. We therefore adopt $\Delta\mathrm{DM}=1.1~\mathrm{pc\,cm^{-3}}$ as a physically motivated absolute-error threshold for evaluating model accuracy, appropriately scaled to the observing band.

The Gaussian peak time $t_{0}$ for each frequency channel was determined by the frequency-dependent arrival time of the burst,
\begin{align}
t_0(\nu) = a\,\mathrm{DM}\!\left(\frac{1}{\nu^2}-\frac{1}{\nu_\mathrm{ref}^2}\right)
- t_\mathrm{w}\!\left(\frac{\nu-\nu_\mathrm{max}}{A_0\nu_\mathrm{ref}}\right),
\label{Eq:arrival}
\end{align}
where the first term describes the dispersion delay introduced by propagation through the ionized medium, and the second term introduces the sub-burst frequency drift predicted by the TRDM \citep{rajabi2020,chamma2021,jahns2022,chamma2023,Brown2024}. We adopt $A_0 = 0.1$, consistent with previous studies. The peak time $t_{0}(\nu)$ thus represents the arrival time of the FRB signal at each frequency channel compared to the arrival time in the reference channel. By assigning this value as the Gaussian center for every channel, the collection of Gaussians across the band naturally reproduces the characteristic dispersion sweep observed in real FRB dynamic spectra.

Scattering effects, which both broaden the FRB pulse and slightly delay its peak, were incorporated by convolving the intrinsic Gaussian profile with a one-sided exponential kernel,
\begin{align}
h(t) = \frac{1}{\tau_\mathrm{s}}\exp\!\left(-\frac{t}{\tau_\mathrm{s}}\right),
\label{eq:h(t)}
\end{align}
where the convolution produces an asymmetric pulse with an exponential tail. The scattering timescale, $\tau_\mathrm{s}$, varies with frequency according to
\begin{align}
\tau_\mathrm{s} = \tau_0\!\left(\frac{\nu_\mathrm{ref}}{\nu}\right)^4,
\label{eq:tau_sc}
\end{align}
consistent with the expected $\nu^{-4}$ dependence for scattering in a Kolmogorov-type turbulent medium. The reference scattering timescale, $\tau_0$, was drawn from a log-uniform distribution to ensure positive values and to reproduce the observed diversity in scattering strengths, where most bursts exhibit moderate scattering while a smaller fraction display long exponential tails. This treatment yields the characteristic asymmetric, exponentially broadened profiles commonly observed in scattered FRB signals.

To further emulate propagation through an inhomogeneous medium, scintillation was introduced as an amplitude modulation across the observing band:
\begin{align}
A(\nu) = \cos\!\left[2\pi N_\mathrm{scint}\!\left(\frac{\nu_\mathrm{ref}}{\nu}\right)^2 + \phi_\mathrm{scint}\right],
\label{eq:scintillation}
\end{align}
where the strength of the scintillation is controlled by the parameter $N_\mathrm{scint}$, which was drawn from a log-uniform distribution between 0.001 and 7. This distribution captures a wide range of scintillation intensities while favouring milder effects that are more frequently observed in nature. The frequency dependence, $\left(\nu_\mathrm{ref}/\nu\right)^2$, reflects the stronger scintillation observed at lower frequencies, whereas the random phase $\phi_\mathrm{scint}$, uniformly distributed between 0 and $2\pi$, introduces variability in the interference pattern across different realizations. This treatment reproduces the complex, frequency-dependent amplitude modulations characteristic of real FRB spectra.

After incorporating these propagation effects, each simulated pulse
was injected into a zero-mean, unit-variance noise background, such
that its signal-to-noise ratio was defined relative to the noise standard
deviation. The overall pulse intensity was then rescaled by a fluence
factor $F$, sampled using an inverse-uniform scheme equivalent to a
Euclidean distribution $p(F) \propto F^{-5/2}$, corresponding to an
approximately Euclidean source population. This choice is consistent
with standard assumptions in FRB population studies and is widely
adopted in simulations of transient surveys (e.g., \citealt{connor2018applying};
\citealt{petroff2019fast}). Such a distribution naturally produces a broad
range of signal-to-noise ratios, with many faint events and fewer
bright bursts. A spectral index $\gamma$, sampled uniformly from the
interval $[-4, 4]$, was applied to introduce frequency-dependent
amplitude variations according to $F_\nu \propto \nu^{\gamma}$. The
resulting frequency–time profiles constitute the complete synthetic
dynamic spectra used for training and evaluation (see Figure~1).

\begin{figure}
    \centering
    \includegraphics[width=\columnwidth]{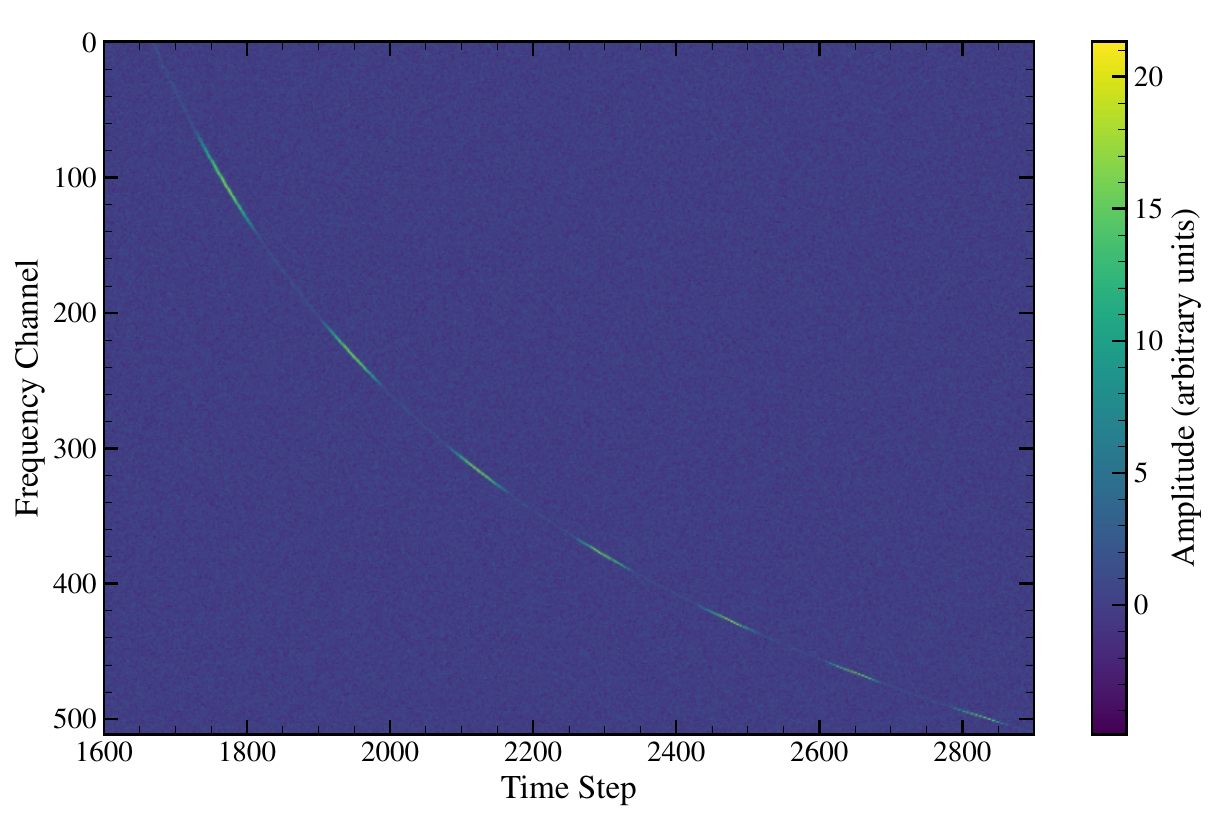}
    \caption{Example of a simulated FRB dynamic spectrum with $\mathrm{DM}=105.362~\mathrm{pc\,cm^{-3}}$.}
    \label{fig:exampledata}
\end{figure}

A complete list of the parameters, equations, and sampling distributions used in these simulations is provided in Appendix~\ref{app:parameters}.

\section{Model training}\label{sec:model_training}

The design and training procedure for our deep-learning models used in DM estimation are outlined below.
Model training was performed on the Narval, Nibi, and Rorqual clusters provided by the Digital Research Alliance of Canada, 
equipped with NVIDIA A100 and H100 GPUs, 128~GB of RAM, and high-end CPUs (AMD EPYC series or Intel Xeon 6972P).
The same hardware configuration was used for both training and validation.

Model performance was evaluated using three standard regression metrics: the Mean Squared Error (MSE),
Root Mean Squared Error (RMSE), and Mean Absolute Error (MAE), defined respectively as
$\mathrm{MSE} = n^{-1}\sum_{i=1}^{n}(y_i-\hat{y}_i)^2$,
$\mathrm{RMSE} = \sqrt{n^{-1}\sum_{i=1}^{n}(y_i-\hat{y}_i)^2}$, and
$\mathrm{MAE} = n^{-1}\sum_{i=1}^{n}|y_i-\hat{y}_i|$,
where $y_i$ and $\hat{y}_i$ are the true and predicted dispersion measures, and $n$ is the number
of samples. MSE quantifies the mean of squared deviations and was also used as the training
loss function for the hybrid CNN--LSTM model owing to its sensitivity to large errors.
RMSE expresses the same quantity in the original units of DM, providing an interpretable measure
of typical prediction error, while MAE reflects the average absolute deviation. MAE was also used as the optimization objective for the baseline CNN and ResNet-50 models.
Together, these metrics provide a balanced and complementary assessment of model accuracy and
robustness in estimating dispersion measures from FRB data.
Additional performance indicators, including relative and global accuracy metrics, are introduced
in Section~\ref{sec:results} to facilitate comparison among the trained models.

\subsection{Baseline Convolutional Neural Network (CNN) Model}
\label{sec:baseline_CNN}

The baseline architecture follows the original design of a convolutional neural network (CNN; \citealt{lecun1998gradient}) developed by \citet{Rajabi2024}, implemented here for DM estimation from frequency–time dynamic spectra. The input dynamic spectra were resized to $224 \times 224$ pixels to ensure compatibility with the network architecture. As described below, the final model comprises seven convolutional layers followed by a sequence of fully connected (dense) layers, as summarised in Table~\ref{tab:cnn_layers}. Each convolutional layer applies a set of filters (kernels) to detect local spectro–temporal features and hierarchically capture complex patterns in the input data. To stabilise training and accelerate convergence, batch normalisation \citep{DBLP:journals/corr/IoffeS15} is applied after each convolutional layer, followed by a Rectified Linear Unit (ReLU) activation \citep{DBLP:conf/icml/NairH10} to introduce non-linearity. Max-pooling operations with $2\times2$ kernels are used between most convolutional layers (except the third and sixth) to reduce the spatial resolution of feature maps while retaining the most significant features, thereby improving computational efficiency and generalisation.

The convolutional layers progressively increase in depth, starting with 32 filters in the first layer and doubling in number through successive layers, culminating in 2048 filters in the seventh (final) convolutional layer, following the same principle as VGGNet (Visual Geometry Group; \citealt{simonyan2014very}). This design follows standard CNN practice, where increasing the number of filters with depth enables the network to capture progressively more complex representations, from simple spectral features in early layers to more complex temporal correlations in deeper ones. Following the convolutional block, a series of fully connected (FC) layers integrates the extracted features to produce a single scalar output corresponding to the predicted DM. The dense section reads the flattened activations (100,352 elements) from the convolutional block, encodes them into 16,384 outputs, and then halves in size through successive layers until reaching the output neuron. The complete layer-by-layer configuration is summarised in Table~\ref{tab:cnn_layers} (adapted from \citealt{Rajabi2024}), where the numbers of inputs and outputs ($a{\to}b$) indicate the number of feature maps entering and leaving each convolutional or dense layer.

\begin{table}
\centering
\caption{Architecture of the baseline CNN model for DM estimation. 
\textit{Architecture summary:} 
The CNN comprises seven convolutional layers (Conv2D-1 to Conv2D-7) with increasing feature depth, each followed by batch normalisation and ReLU activation, and five max-pooling layers that progressively reduce the spatial dimensions of the feature maps. The flattened output is passed through five fully connected layers (FC-1 to FC-5) that map the extracted features to a single scalar output representing the predicted dispersion measure. \textit{Notation:} ``in/out'' = input/output channels; ``k'' = kernel size (height$\times$width); ``s'' = stride length (step size of the convolution); ``p'' = zero-padding width (e.g. $p=1$ adds one zero on each side, preserving output size); ``BN'' = batch normalisation; ``FC'' = fully connected layer.}
\label{tab:cnn_layers}
\renewcommand{\arraystretch}{1.08}
\setlength{\tabcolsep}{4pt}
\footnotesize
\begin{tabular}{@{}p{0.33\columnwidth} p{0.65\columnwidth}@{}}
\hline
\textbf{Layer (Type)} & \textbf{Configuration} \\
\hline
Conv2D-1 + BN + ReLU & in/out: $1{\to}32$; k = $3{\times}3$; s = 1; p = 1 \\
MaxPool2d-1          & k = $2{\times}2$; s = 2 \\[3pt]
Conv2D-2 + BN + ReLU & in/out: $32{\to}64$; k = $3{\times}3$; s = 1; p = 1 \\
MaxPool2d-2          & k = $2{\times}2$; s = 2 \\[3pt]
Conv2D-3 + BN + ReLU & in/out: $64{\to}128$; k = $3{\times}3$; s = 1; p = 1 \\[3pt]
Conv2D-4 + BN + ReLU & in/out: $128{\to}256$; k = $3{\times}3$; s = 1; p = 1 \\
MaxPool2d-3          & k = $2{\times}2$; s = 2 \\[3pt]
Conv2D-5 + BN + ReLU & in/out: $256{\to}512$; k = $3{\times}3$; s = 1; p = 1 \\
MaxPool2d-4          & k = $2{\times}2$; s = 2 \\[3pt]
Conv2D-6 + BN + ReLU & in/out: $512{\to}1024$; k = $3{\times}3$; s = 1; p = 1 \\[3pt]
Conv2D-7 + BN + ReLU & in/out: $1024{\to}2048$; k = $3{\times}3$; s = 1; p = 1 \\
MaxPool2d-5          & k = $2{\times}2$; s = 2 \\[3pt]
\cline{1-2}
Flatten              & $7{\times}7{\times}2048 = 100{,}352$ features \\[3pt]
\cline{1-2}
FC (1--5)            & 100,352 $\to$ 16,384 $\to$ 8192 $\to$ 4096 $\to$ 2048 $\to$ 1 \\
\hline
\end{tabular}
\end{table}

During hyper-parameter tuning, we explored architectures ranging from four to seven convolutional layers and varied the widths of the dense layers to achieve an optimal balance between model complexity and overfitting. Batch normalisation significantly improved convergence stability by reducing internal covariate shift \citep{DBLP:journals/corr/IoffeS15}, enabling the use of higher learning rates without divergence. Although dropout is a common regularisation technique \citep{srivastava2014dropout}, tests with dropout probabilities between 0 and 0.2 showed that random unit removal degraded performance, since strong spatial dependencies exist between neighbouring pixels in the input data \citep{zhang2018fast}. Consequently, dropout layers were omitted.

The model minimises the MAE loss function, which is well suited to continuous regression tasks such as DM estimation. We used the Adam algorithm \citep{adam2014method} as the optimiser, which adapts learning rates for individual parameters based on mini-batch updates and is well suited for large datasets in deep learning. Learning rates in the range $[10^{-7},10^{-2}]$ were tested, with the smallest value ($10^{-7}$) yielding the smoothest convergence and most accurate results. Larger learning rates led to divergence from the optimal solution, whereas smaller rates allowed for gradual and stable parameter updates.

This baseline CNN serves as the reference model against which the more advanced architectures, ResNet-50 (Section~\ref{sec:resnet50}) and the CNN–LSTM hybrid (Section~\ref{sec:HybridModel}), are evaluated in subsequent sections.


\subsection{ResNet-50}
\label{sec:resnet50}

ResNet-50, a deep convolutional neural network comprising 50 layers \citep{he2016deep}, was fine-tuned to adapt its pre-trained features for the regression task of DM estimation. The residual connections of the network, or skip connections, enable efficient training of deep architectures by alleviating the problem of vanishing gradients \citep{he2016deep}. To repurpose the model for continuous DM prediction, its original classification head, consisting of 1,000 neurons corresponding to ImageNet classes, was replaced with a single linear output neuron. The earlier convolutional layers, which capture general image features such as edges and textures, were kept frozen to retain their pre-trained weights, while the final two convolutional stages (Layers~4 and~5) were unfrozen and trained to extract task-specific patterns from the FRB data. We used the AdamW optimizer with weight decay, which provides an effective and stable form of regularization by directly penalizing large weights, thereby reducing overfitting and preventing the model from overreacting to specific features. We also employed a learning-rate scheduler that reduces the learning rate by a factor of 0.4 every ten training epochs, enabling stable convergence over extended training and allowing further exploration of the model’s learning capacity.

To ensure compatibility with the pre-trained model, the FRB data were pre-processed as follows. During training, an additional noise background was first added to the input spectrograms, where each element was sampled from a uniform distribution in the range $[0,1]$. This augmentation introduces variability in the background and improves robustness to noise in real observations. Each image was then normalised using min--max scaling to the range [0,\,1], resized to 224\,$\times$\,224\,pixels, and replicated across three channels to satisfy the RGB input requirements of ResNet-50. The images were subsequently standardised using the ImageNet mean and standard deviation values ([0.485,\,0.456,\,0.406] and [0.229,\,0.224,\,0.225], respectively). The model was trained using the AdamW optimiser \citep{adam2014method} with a learning rate of $10^{-4}$ and the MAE loss function, which is well suited to continuous regression tasks. Training was performed for 150\,epochs, and model performance was evaluated using MAE and RMSE metrics.

Through this fine-tuning procedure, ResNet-50 was adapted for DM estimation, leveraging its residual learning framework to capture high-level spectro–temporal features and achieve accurate and robust predictions.


\subsection{CNN--LSTM Hybrid Model}
\label{sec:HybridModel}

To improve computational efficiency while preserving accuracy, we developed a hybrid convolutional neural network–long short-term memory (CNN–LSTM) architecture. This design combines convolutional layers for spectro--temporal feature extraction with recurrent layers for temporal modelling, enabling the network to learn both the spectro–temporal and sequential characteristics of FRB signals. The hybrid approach substantially reduces the number of trainable parameters relative to the baseline CNN, shortening training time without compromising predictive performance. Such a model has been shown to be robust in principle, as demonstrated by earlier work on radio transient classification \citep{czech2018cnn}.

The recurrent component utilises Long Short-Term Memory (LSTM) units \citep{hochreiter1997long}, which mitigate the vanishing-gradient problem of standard recurrent neural networks by introducing gating mechanisms that control information flow through time. Each LSTM cell maintains two information pathways: the \textit{cell state}, which stores long-term memory, and the \textit{hidden state}, which captures short-term dynamics. This dual mechanism enables the network to retain context across the entire time steps of the input waterfall plots, effectively capturing both transient spectral variations and long-term temporal trends. 

\begin{figure}
    \centering
    \includegraphics[width=0.95\columnwidth, trim= 0cm 0.8cm 0cm 
    0.8cm]{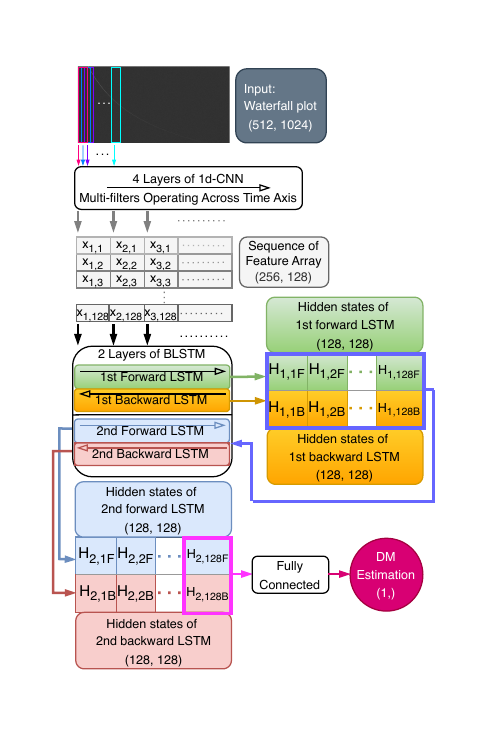}
    \caption{Activation flow through the hybrid CNN--LSTM model during DM estimation. The model input is a waterfall plot of size $512 \times 1024$, representing frequency (512 channels) and time (1024 samples). The CNN front end consists of four one-dimensional convolutional layers that apply multiple filters along the time axis across all frequency channels, producing a compact sequence of feature maps. The resulting representation has size $(256, 128)$. Here, 256 denotes the number of learned feature channels, while 128 corresponds to the reduced temporal dimension obtained from the original 1024 time samples through successive pooling operations. This feature sequence is passed to the recurrent back end, composed of two bidirectional LSTM (BLSTM) layers. In each BLSTM layer, the feature sequence is processed in both \textit{forward} and \textit{backward} temporal directions, generating hidden states $H_\mathrm{F}$ and $H_\mathrm{B}$ with 128 units each. The hidden states from both directions are concatenated along the feature dimension at each time step and passed to the next BLSTM layer, enabling the network to capture dependencies in both temporal directions. The fully connected regression head receives input from the final concatenated hidden states of the second BLSTM layer and produces a single scalar output corresponding to the estimated DM. This structure enables the model to integrate local spectro--temporal features extracted by the CNN with long-range temporal context learned by the bidirectional LSTM layers.}  \label{fig:LSTMInfomation}
\end{figure}

The hybrid network operates sequentially, with the convolutional and recurrent components 
working in tandem to extract and integrate spectro–temporal features. As shown in 
Figure~\ref{fig:LSTMInfomation}, the CNN front end processes the frequency–time waterfall 
plots using one-dimensional convolutions applied along the temporal axis of each frequency 
channel. Local rectangular kernels ($k = 1{\times}5$, depth equal to the number of frequency 
channels) detect short-term variations in spectral intensity, while deeper kernels ($k = 
1{\times}3$) span multiple feature channels to capture inter-channel correlations. The choice 
of kernel sizes reflects a balance between capturing short-duration temporal structures and 
maintaining computational efficiency, as validated through empirical testing during model 
development. The convolutional layers progressively increase in depth from 32 to 256 filters, 
capturing increasingly abstract spectro–temporal features. Each convolutional layer is 
equipped with a batch normalisation layer to stabilise feature distributions during training. 
Max-pooling layers follow most convolutional layers (except the third layer) to downsample 
the temporal sequence, reducing computational load while retaining dominant features. The 
placement of max-pooling layers was determined empirically to balance dimensionality 
reduction and feature preservation, as additional pooling in deeper layers did not improve 
validation performance and led to a loss of temporal resolution, resulting in no net gain in 
model accuracy. The resulting compact feature sequence is then passed to the recurrent back 
end.

The output sequence from the CNN encoder is processed by two stacked bidirectional LSTM (BLSTM) layers, each with 128 hidden units per direction. These layers read the feature sequence in both forward and backward temporal directions, enabling the network to learn long-range dependencies while preserving local context. The hidden states from both directions are concatenated after sequential aggregation and fed into a fully connected regression head. The flow of activations through these stages is illustrated in Figure~\ref{fig:LSTMInfomation}, where the hidden states from the second BLSTM layer are passed to two dense layers that output the predicted DM. The detailed configuration of the hybrid architecture, including filter counts, kernel sizes, and recurrent-layer dimensions, is summarised in Table~\ref{tab:LSTMparameters}.

Training employed the Adam optimizer with an initial learning rate of $10^{-4}$, reduced by a factor of 0.4 every ten epochs via a scheduler to ensure stable convergence. The MSE loss function was selected for its sensitivity to large deviations, which helps suppress
outlier predictions and stabilise the learning process. Dropout layers were omitted, as no
evidence of overfitting was observed in the validation results. The model was trained for
150 epochs with a batch size of 16. Each epoch required approximately 17 minutes on an
NVIDIA A100 GPU, and the final trained model had a compact memory footprint of only 2.4~MB.

\begin{table}
\centering
\caption{Architecture of the hybrid CNN--LSTM model. 
The CNN front end (Conv1--Conv4) performs temporal convolutions applied along the time axis of the frequency--time representation (kernels $1{\times}5$, then $1{\times}3$) to extract local spectro--temporal patterns. 
Max-pooling is applied after selected convolutional layers (Conv1, Conv2, and Conv4) to reduce the temporal dimension while preserving dominant features. 
The recurrent back end (BLSTM1--2) processes the resulting feature sequence in both forward and backward directions to learn short- and long-range temporal dependencies. 
The regression head (FC-1, FC-2) takes the concatenated hidden states at the last time step and outputs the predicted DM. 
Notation: ``in/out'' = input/output channels; ``k'' = kernel size (height$\times$width); ``s'' = stride length; ``p'' = zero-padding width; ``BN'' = batch normalisation; ``FC'' = fully connected layer.}
\renewcommand{\arraystretch}{1.06}
\setlength{\tabcolsep}{3pt}
\footnotesize
\begin{tabular}{@{}p{0.30\columnwidth} p{0.66\columnwidth}@{}}
\hline
\textbf{Layer (Type)} & \textbf{Configuration} \\
\hline
Conv1 + ReLU + BN  & in/out: $512{\to}32$; k = $1{\times}5$; s = 1; p = 2 \\
MaxPool1          & window 2; s = 2 \\
Conv2 + ReLU + BN & in/out: $32{\to}64$; k = $1{\times}3$; s = 1; p = 1 \\
MaxPool2          & window 2; s = 2 \\
Conv3 + ReLU + BN & in/out: $64{\to}128$; k = $1{\times}3$; s = 1; p = 1 \\
Conv4 + ReLU + BN & in/out: $128{\to}256$; k = $1{\times}3$; s = 1; p = 1 \\
MaxPool3          &window 2; s = 2 \\
BLSTM-1           & input 256; hidden units: 128 (forward) + 128 (backward) \\
BLSTM-2           & input 256; hidden units: 128 (forward) + 128 (backward) \\
FC-1              & input 256 (hidden states, last time step); 128 outputs \\
FC-2              & input 128; 1 output \\
\hline
\label{tab:LSTMparameters}
\end{tabular}
\end{table}

\section{Results and Discussion}\label{sec:results}

\subsection{Training and Validation Performance}\label{sec:training-performance}

Each of the three deep-learning architectures, the baseline CNN, the fine-tuned ResNet-50, and the hybrid CNN–LSTM, was trained under identical six-fold cross-validation. Detailed fold-by-fold results are presented in Appendix~\ref{app:crossvalidation}, while this section
summarises the performance averaged across all folds. The baseline CNN and ResNet-50 models
were optimised using the MAE loss function, whereas the hybrid CNN--LSTM
model employed the MSE loss function. All models were validated using
MAE and RMSE as absolute-error metrics, along with the mean error
percentage (MErrP), defined as
$\mathrm{MErrP} = n^{-1}\sum_{i=1}^{n}\frac{|y_i-\hat{y}_i|}{y_i}\times100$,
where $y_i$ and $\hat{y}_i$ are the true and predicted dispersion measures, respectively, and $n$ is the number
of samples. MErrP is reported in per cent throughout.

\begin{figure*}
    \centering
    \includegraphics[width=\textwidth]{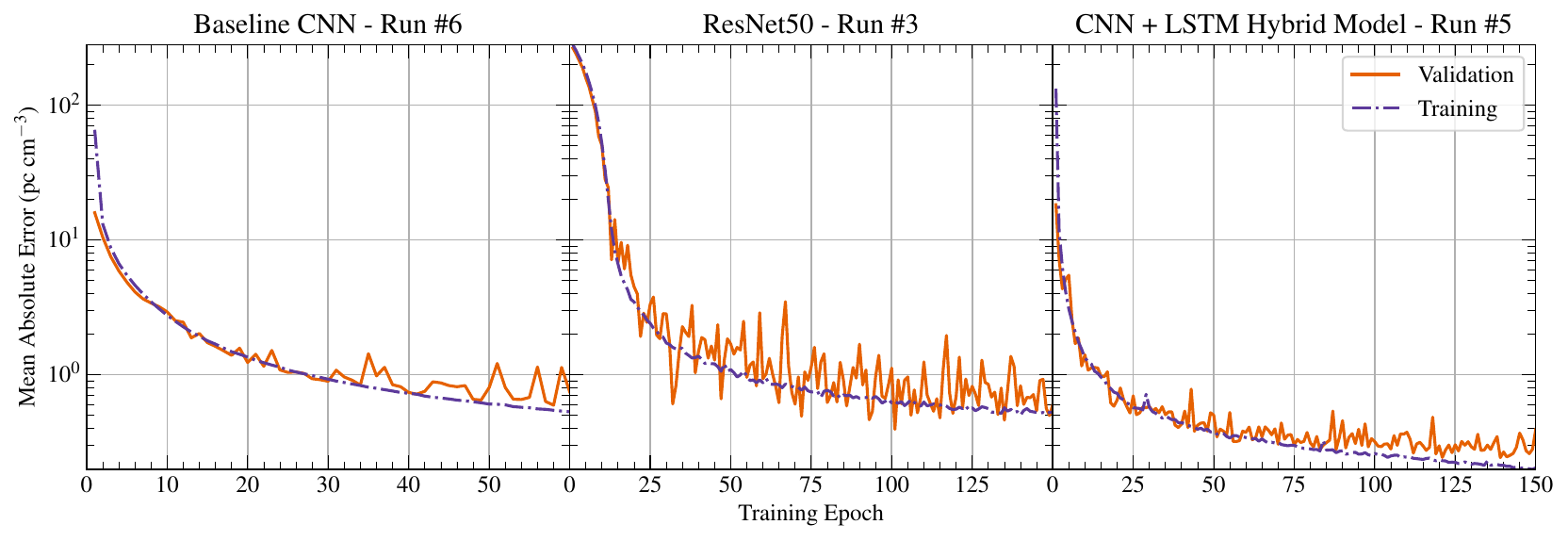}
    \caption{Learning curves for the best-performing fold of each model. The panels show the mean absolute error (MAE) of the training and validation datasets as a function of epoch number. 
    The baseline CNN was trained for 60~epochs, whereas the ResNet-50 and hybrid CNN–LSTM models were trained for 150~epochs, taking advantage of shorter epoch durations and smaller memory footprints.}
    \label{fig:learning process}
\end{figure*}

The learning curves in Figure~\ref{fig:learning process} trace the evolution of the models’ MAE
during training, illustrating their convergence behaviour across epochs. All models exhibit stable convergence, although their learning behaviours differ. 
The CNN and hybrid models show a gradual and steady decline in loss, whereas ResNet-50 displays greater fluctuation during optimisation. 
The hybrid CNN–LSTM converges most rapidly, reaching an MAE of approximately 1~pc\,cm$^{-3}$ after only $\sim$15~epochs, followed by the CNN after $\sim$28~epochs, while ResNet-50 requires $\sim$55~epochs. 
Despite its slower initial learning rate, ResNet-50 demonstrates strong generalisation due to its residual connections. 
The hybrid model displays the smoothest and most consistent validation behaviour, combining rapid convergence with the lowest final MAE.

The baseline CNN’s accuracy is ultimately limited by its architectural simplicity.
Improving performance by increasing convolutional depth alone would require substantially
deeper networks, leading to increased training time and computational cost.
In contrast, incorporating residual or recurrent components can provide a more efficient
alternative, enabling faster training and reduced computational cost for a given level
of performance.

ResNet-50 mitigates gradient degradation through skip connections, enabling deeper feature
extraction, while the hybrid CNN--LSTM further improves performance by incorporating
temporal correlations directly into its learning process.

The average results from the six-fold cross-validation are listed in Table~\ref{tab:cv_summary}. The results show a clear progression across the three architectures, with the baseline CNN providing a reference level of accuracy, ResNet-50 yielding improved performance, and the hybrid CNN–LSTM achieving the lowest errors overall. A similar trend is observed in MErrP, which also decreases steadily from the CNN to the hybrid model. 
The close agreement between training and validation metrics across all architectures confirms that none of the models show signs of overfitting. 
Overall, the hybrid CNN–LSTM provides the most accurate and stable predictions among the models considered.
\begin{table}
\centering
\caption{Six-fold cross-validation \emph{averages} (means across folds) of training and validation metrics for each model. The unit for MAE and RMSE are~pc\,cm$^{-3}$. Per-fold values are reported in Appendix~\ref{app:crossvalidation}.}
\label{tab:cv_summary}
\setlength{\tabcolsep}{5.2pt} 
\renewcommand{\arraystretch}{1.08}
\footnotesize
\begin{tabular}{l@{\hspace{10pt}}ccc}
\hline
\multicolumn{4}{c}{\textbf{Training (averages across 6 folds)}} \\
\hline
\textbf{Model} & \textbf{MAE} & \textbf{RMSE} & \textbf{MErrP} \\
\hline
CNN              & 0.5754 & 1.1945 & 4.00 \\
ResNet-50        & 0.5442 & 1.4882 & 1.28 \\
Hybrid CNN--LSTM & 0.2162 & 0.2915 & 0.84 \\
\hline
\multicolumn{4}{c}{\textbf{Validation (averages across 6 folds)}} \\
\hline
\textbf{Model} & \textbf{MAE} & \textbf{RMSE} & \textbf{MErrP} \\
\hline
CNN              & 0.6723 & 1.4469 & 3.84 \\
ResNet-50        & 0.4802 & 0.7294 & 2.46\\
Hybrid CNN--LSTM & 0.2530 & 0.6070 & 0.86 \\
\hline
\end{tabular}
\end{table}

\subsection{Test-set Results}\label{sec:test-performance}

Building on the cross-validation results, we next evaluated each model’s generalisation on an
independent test dataset drawn from the same DM distribution but excluded from the training
and fine-tuning phases. Table~\ref{tab:test_summary} summarises the averaged test-set performance
across all six folds, while detailed per-fold results are provided in Appendix~\ref{app:testresults}.
The reported metrics include the MAE, RMSE, MErrP, and two complementary accuracy indicators: the global effective rate (GER) and the global deviation rate (GDR). The GDR quantifies the fraction of predictions with absolute errors below the adopted threshold of $1.1~\mathrm{pc\,cm^{-3}}$, while the GER measures the fraction with relative errors below $1$~per~cent. Together, these metrics
provide a comprehensive view of predictive precision and robustness. As shown in Table~\ref{tab:test_summary}, all models follow the same general trend observed during cross-validation,
with prediction errors decreasing and overall accuracy improving as architectural depth increases and explicit temporal modelling is subsequently introduced.

\begin{figure*}
    \centering
    \includegraphics[width=\textwidth]{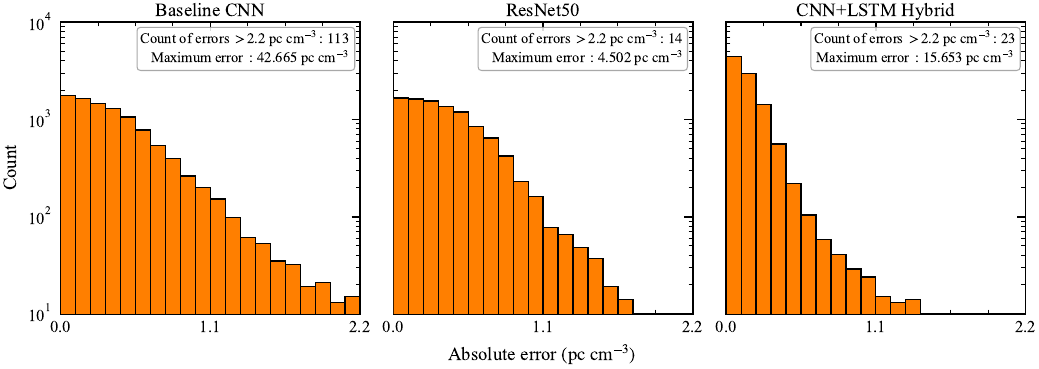}
    \caption{Each panel shows the histogram of absolute errors $|\Delta \mathrm{DM}|$ (in pc\,cm$^{-3}$) for 10,000 test samples; the vertical axis gives the count of predictions per bin on a logarithmic scale. The baseline CNN exhibits the broadest distribution (113 outliers with $|\Delta \mathrm{DM}| > 2.2$\,pc\,cm$^{-3}$; maximum $\sim 42$\,pc\,cm$^{-3}$), ResNet-50 is narrower (14 outliers; maximum $\sim 4.5$\,pc\,cm$^{-3}$), and the hybrid CNN--LSTM shows the highest concentration near small errors but a slightly larger number of outliers and a higher maximum error than ResNet-50 (23 outliers; maximum $\sim 15.7$\,pc\,cm$^{-3}$).}
    \label{fig:ErrorDistributionhistogram}
\end{figure*}

Figure~\ref{fig:ErrorDistributionhistogram} illustrates the average results through histograms
of absolute DM-prediction errors for the three models.
The horizontal axis represents the absolute error $|\Delta \mathrm{DM}|$ (in pc\,cm$^{-3}$) for the test set, obtained by excluding the highest and lowest predictions for each
sample across the six folds, and the vertical axis shows the number of test samples in each
error bin on a logarithmic scale.
The baseline CNN exhibits the broadest distribution, with a gradual decline toward higher
errors and a small but noticeable tail extending to a maximum error of
$42.7~\mathrm{pc\,cm^{-3}}$ (not shown in the plot). It also produces the largest number of extreme outliers (113 points out of 10,000 test samples),
defined by $|\Delta \mathrm{DM}|>2.2$~pc\,cm$^{-3}$.
The ResNet-50 histogram is narrower than the baseline CNN, with fewer large-error events (14 outliers)
and a lower maximum error of $4.5~\mathrm{pc\,cm^{-3}}$.
The hybrid CNN--LSTM displays the steepest fall-off, with the highest concentration of
predictions near the low-error bins and a maximum error of
$15.7~\mathrm{pc\,cm^{-3}}$.
Although the hybrid model exhibits more outliers than ResNet-50 (23 versus 14) and a larger
maximum error ($15.7~\mathrm{pc\,cm^{-3}}$ versus $4.5~\mathrm{pc\,cm^{-3}}$), its
substantially higher concentration of very low-error predictions results in smaller MAE and
RMSE values. These progressively sharper peaks and the increasing concentration of low-error predictions correspond to the decreasing MAE listed in Table~\ref{tab:test_summary}. This indicates that the improvement in performance is driven primarily by a higher density of very accurate predictions, despite small variations in the number of outliers.

Quantitatively, the baseline CNN achieves a mean MAE of 0.6690 and RMSE of 1.1045, while ResNet-50 improves these to 0.4802 and 0.6253, representing reductions of approximately 28 per cent and 43 per cent, respectively, relative to the baseline. The hybrid CNN–LSTM achieves the highest precision, with a mean MAE of 0.2542, RMSE of 0.6429, and an average MErrP of 0.26~per cent. Under a 1~per cent relative-error threshold, the hybrid model correctly predicts 96.7~per cent of test samples, compared with approximately 92~per cent for both the ResNet-50 and CNN models. 
Moreover, 98.06~per cent of the hybrid model’s predictions fall below the 1.1~pc\,cm$^{-3}$ absolute-error limit (GDR), confirming its robustness across the entire test range.

\begin{table}
\centering
\caption{Six-fold test-set \emph{averages} (means across folds) for each model. 
Metrics include the mean absolute error (MAE), root mean squared error (RMSE), mean error percentage (MErrP), and the global effective and deviation rates (GER, GDR). 
Per-fold results are provided in Appendix~\ref{app:testresults}.}
\label{tab:test_summary}
\renewcommand{\arraystretch}{1.09}
\footnotesize
\resizebox{\columnwidth}{!}{%
\begin{tabular}{lccccc}
\hline
\textbf{Model} & \textbf{MAE} & \textbf{RMSE} & \textbf{MErrP} & \textbf{GER (\%)} & \textbf{GDR (\%)} \\
\hline
CNN              & 0.6690 & 1.1045 & 0.77 & 92.23 & 83.42 \\
ResNet-50        & 0.4802 & 0.6253 & 0.54 & 92.34 & 93.13 \\
Hybrid CNN--LSTM & 0.2542 & 0.6429 & 0.26 & 96.68 & 98.06 \\
\hline
\end{tabular}%
}
\end{table}


\subsection{Comparative Error Distributions}\label{sec:error-distribution}

To gain deeper insight into the predictive behaviour of the three architectures, we examined the distribution and structure of their prediction errors across the full DM range. 
For this analysis, six independent test runs were performed, and for each model the predictions were averaged after excluding the highest and lowest values to reduce the influence of outliers. 
We then compared these averaged predictions with the ground-truth DMs to compute the overall absolute error.

\begin{figure*}
    \centering
    \includegraphics[width=\textwidth]{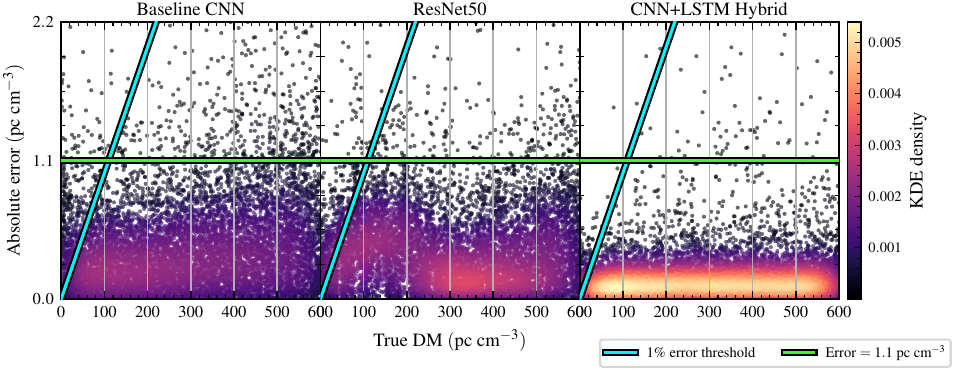}
    \caption{Absolute error as a function of the true DM for the three models. Each scatter plot shows 10,000 test samples, with points color-coded by their kernel density estimate (KDE) to highlight regions of high prediction density. The colour scale represents the estimated density of the point distribution, used to visualise regions of higher data concentration rather than an explicit probability density function. Two solid lines in green and cyan denote the adopted error thresholds: an absolute error of 1.1~pc\,cm$^{-3}$ and a relative error of 1\%, respectively. The hybrid CNN--LSTM exhibits the most symmetric and tightly clustered distribution across the full DM range, demonstrating uniform accuracy and minimal systematic bias.}
    \label{fig:ErrorDistributionThreshold}
\end{figure*}

We visualise the absolute-error distribution as a function of the ground-truth DM for all three models using a kernel density estimation (KDE) in Figure~\ref{fig:ErrorDistributionThreshold}. The KDE is computed using a Gaussian kernel, implemented with the \texttt{scipy.stats} library. This highlights how the magnitude and spread of residuals vary across the 0–600~pc\,cm$^{-3}$ DM range. 
These diagnostics complement the statistical metrics in Table~\ref{tab:test_summary} by revealing how prediction precision and robustness evolve with DM, rather than summarizing them in a single scalar value.

Across all models, errors remain predominantly concentrated within the illustrative thresholds of 1.1~pc\,cm$^{-3}$ (absolute) and 1~per~cent (relative), indicated by the solid green and cyan lines. 
The hybrid CNN–LSTM model maintains the most symmetric and tightly clustered error distribution throughout the DM range, demonstrating both uniform accuracy and minimal systematic bias. 
By contrast, the baseline CNN displays a gradual broadening of scatter toward higher DMs, where the density of low-error points decreases, suggesting a decline in regression precision at the high end of the range. 
The ResNet-50 model performs more uniformly, though a mild concentration of larger residuals appears below $\mathrm{DM}\lesssim200$~pc\,cm$^{-3}$, likely reflecting sensitivity to pre-trained convolutional weights that remained partially frozen during fine-tuning.

The local KDE in Figure~\ref{fig:ErrorDistributionThreshold} further illustrate that, while ResNet-50 yields fewer large-error events than the CNN, the hybrid CNN–LSTM achieves a much higher concentration of very low-error predictions across all DMs. 
This behaviour aligns with the lower MAE and RMSE values reported in Table~\ref{tab:test_summary}, confirming that the hybrid architecture captures the underlying spectro–temporal dependencies most effectively.

These results are consistent with the GER and GDR statistics, indicating that deeper and temporally aware models primarily improve absolute accuracy, while significant gains in fractional accuracy emerge only for the hybrid architecture.

Although some isolated outliers persist for the hybrid CNN–LSTM, mainly corresponding to rare test samples where the MSE loss amplifies occasional mispredictions, these constitute about 0.2~per cent of the test set and have a negligible impact on the global metrics.

In summary, the comparative error analysis demonstrates that model improvements are not confined to mean statistics but extend uniformly across the DM domain. 
Transitioning from the baseline CNN to ResNet-50 reduces both the number and magnitude of outliers, while adding the recurrent component in the hybrid CNN–LSTM further minimizes residual variance and asymmetry. 
This confirms that deeper and temporally aware architectures attain higher predictive precision by more accurately capturing the intrinsic correlations between frequency and time in the FRB dynamic spectra.


\subsection{Computational Efficiency and Scalability}\label{sec:efficiency}

Efficient training and compact model storage are essential for scaling DM estimation to large FRB datasets and for integration into real-time processing pipelines. Table~\ref{tab:efficiency} summarizes the average training times per epoch and checkpoint sizes for the three architectures, measured on the NVIDIA~A100~GPU.

\begin{table}
\centering
\caption{Comparison of computational efficiency and storage requirements for the three
architectures. Reported training times per epoch are approximate averages based on runs
performed on NVIDIA~A100 GPUs and may vary depending on shared-resource allocation on the
supercomputing system.}
\label{tab:efficiency}
\setlength{\tabcolsep}{3pt}
\renewcommand{\arraystretch}{1.1}
\begin{tabular*}{\columnwidth}{@{\extracolsep{\fill}}lcc@{}}
\hline
Model & Training time per epoch & Checkpoint size \\
\hline
Baseline CNN & $\sim$4~h with 2 A100 GPU & 6.9~GB \\
ResNet-50 & $\sim$24~min with 1 A100 GPU & 88~MB \\
Hybrid CNN--LSTM & $\sim$17~min with 1 A100 GPU & 2.4~MB \\
\hline
\end{tabular*}
\end{table}

The high parameter count of the CNN baseline results in high computational and storage costs during training, making it less suitable for continuous retraining to adapt to the evolving characteristics of large-scale dynamic spectral data. In contrast, the ResNet-50 leverages pre-trained initialization to dramatically reduce training time and checkpoint size. The hybrid CNN–LSTM is similarly fast but considerably more compact, owing to its use of one-dimensional convolutions and parameter sharing in recurrent layers.

In addition to training efficiency, we evaluated the end-to-end inference performance of each model to assess suitability for real-time applications. Inference times were measured on an NVIDIA~A100 GPU for samples of 100 waterfall plots. Inference time was measured from when the DataLoader begins loading the first batch. This includes disk input, device transfers, time-axis downsampling, specialized preprocessing for each model, and the forward pass, until the model produces the DM prediction for the last batch. The time required for model loading and DataLoader initialization was excluded, as they are considered one-time costs. The three models exhibited comparable inference times, with differences at the sub-second level (7.3--7.6~s per 100 samples).

Each input sample corresponds to a dynamic spectrum of 7340 time bins with a sampling time of $\sim1.67$~ms, yielding a duration of $\sim12.3$~s of telescope data. The effective throughput is therefore obtained by converting the total duration of the processed samples into an equivalent amount of telescope data processed per unit time.

Given this, the measured inference times correspond to processing approximately $161$--$167$~s of telescope data per second of computation. This effective throughput significantly exceeds real-time processing rates and is substantially higher than that of traditional GPU-based dedispersion pipelines (e.g., $\sim$3$\times$ real-time; \citealt{Barsdell2012}). However, the two approaches are not directly comparable, as conventional pipelines perform an explicit search over a large number of DM trials, whereas the present model infers the DM in a single forward pass.

Together with their short training times and compact model checkpoints, these results demonstrate that the ResNet-50 and hybrid CNN--LSTM architectures are promising candidates for scalable, low-latency solutions for real-time DM estimation in large FRB surveys.

\section{Conclusions and Future Work}\label{sec:conclusion}

We have benchmarked three deep-learning architectures, a conventional CNN, a fine-tuned ResNet-50, and a hybrid CNN--LSTM, for quantitative DM estimation in FRBs. Trained and validated on a uniform synthetic dataset, the models were evaluated using consistent metrics to assess their accuracy, convergence, and computational efficiency. The results demonstrate that increasing architectural depth and incorporating temporal modelling lead to clear performance gains, with the hybrid CNN--LSTM providing the most robust overall performance, combining high precision, stable convergence, and low computational cost.

Across all diagnostics (Figures~\ref{fig:learning process}--\ref{fig:ErrorDistributionThreshold}; Tables~\ref{tab:cv_summary}--\ref{tab:efficiency}), the baseline CNN provides a reliable benchmark, while the ResNet-50 and hybrid CNN--LSTM architectures exhibit progressively improved accuracy, scalability, and throughput. The hybrid model reduces the mean absolute error by nearly 60\,per\,cent, relative to the baseline CNN and attains a global deviation rate of $\sim$98\,per\,cent, confirming its robustness across the full DM range. The achieved precision satisfies the tolerance required for spectro--temporal analyses under TRDM \citep{rajabi2020}, ensuring that DM uncertainties remain below levels that affect sub-burst slope measurements.

Traditional FRB dedispersion pipelines rely on real-time incoherent or coherent algorithms that primarily maximize signal-to-noise ratio, enabling rapid detection but often at the expense of fine-scale structural fidelity, particularly for low-S/N or morphologically complex bursts. In contrast, high-precision post-processing techniques such as DM phase \citep{Seymour2019}, DM power \citep{Lin2022}, or SVD-based substructure maximization can achieve sub--pc\,cm$^{-3}$ accuracy for bright, well-resolved events, albeit with increased computational cost and, in many cases, manual intervention. The ML framework developed here is intended as a complementary approach, with the potential to provide stable DM estimates over a broad range of burst properties and, with further development, to operate efficiently at survey scale.

As a further step toward assessing applicability to real observations, we performed preliminary tests of the trained models on a sample of high S/N CHIME/FRB bursts. The results show that the models can recover meaningful dispersion measures and capture relevant spectro--temporal features present in real data. However, when evaluated against the physically motivated accuracy requirement of $\Delta\mathrm{DM} = 1.1~\mathrm{pc\,cm^{-3}}$, the performance remains below the level achieved on simulated data, reflecting the increased complexity of real observations. In particular, a subset of outliers persists, primarily associated with more complex burst morphologies (e.g., multi-component bursts), strong scattering, and residual RFI or background artifacts that are not fully removed during preprocessing. These effects introduce structures that are not fully represented in the simulated training data, leading to degraded performance. These results indicate that, while the proposed framework is effective under controlled conditions, further development is required before it can be reliably applied to real observational pipelines.

Future work will therefore focus on incorporating more realistic noise characteristics, including RFI and background variability, into the training process, as well as improving preprocessing strategies and extending the simulations to better capture the diversity of real burst morphologies. These developments will be essential for enabling reliable application of such models to real observational data and for adapting them to different telescope configurations.

\section*{Acknowledgements}

M.H.’s research is funded through the Natural Sciences and Engineering Research Council of Canada (NSERC) Discovery Grant RGPIN-2024-05242. F.R.'s research is supported by the NSERC Discovery Grant RGPIN-2024-06346. F.R. is grateful for the hospitality of Perimeter Institute where part of this work was carried out. Research at Perimeter Institute is supported in part by the Government of Canada through the Department of Innovation, Science and Economic Development and by the Province of Ontario through the Ministry of
Colleges and Universities. This work was supported by a grant from the Simons Foundation (1034867, Dittrich).
\section*{Data availability}
The data underlying this article will be shared on reasonable request to the corresponding author.



\bibliographystyle{mnras}
\bibliography{references} 



\appendix

\section{Distribution of Dispersion Measure in the Simulated Data}\label{appendix:dm_distribution}

\begin{figure}
    \centering
    \includegraphics[width=\columnwidth]{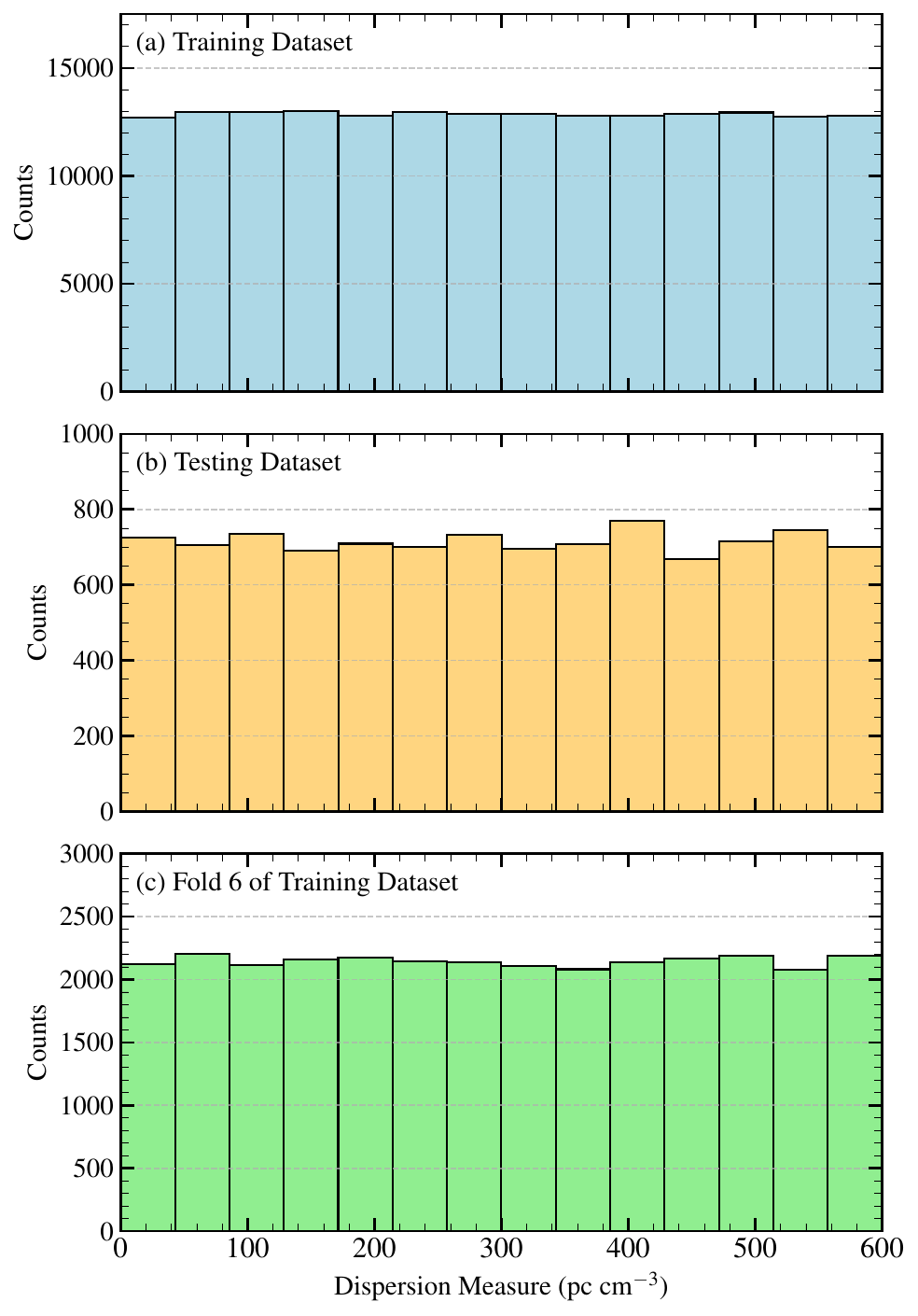}
    \caption{Distribution of dispersion measures (DMs) for three subsets of the synthetic dataset:
    (a) training set, (b) testing set, and (c) fold 6 of the training set as an example. 
    All panels use identical binning for comparison, illustrating the uniform DM coverage across all subsets.}
    \label{fig:dm_histograms}
\end{figure}

The DM distributions for the three subsets of the synthetic data are shown in Figure \ref{fig:dm_histograms}.

\section{Simulation Parameters}\label{app:parameters}

Table~\ref{tab:sim_params} summarises the parameters, governing equations, and sampling distributions used to generate the synthetic FRB dataset described in Section~\ref{sec:data-simulation}. The simulations reproduce key observational and instrumental effects, including dispersion, scattering, scintillation, and finite time and frequency resolution. 

Dispersion measures were uniformly drawn between 0 and 600~pc\,cm$^{-3}$, consistent with the CHIME/FRB-like observing band adopted in this study. A sampling time of $t_\mathrm{samp}=1.67$~ms ensures that the dispersion sweep remains within the simulated time window. Other parameters, such as the intrinsic pulse width, scattering timescale, scintillation strength, and background noise, were drawn from distributions designed to mimic the observed diversity of FRB properties.

\begin{table*}
\centering
\caption{Parameters, governing equations, and sampling distributions used in the FRB simulations.}
\label{tab:sim_params}
\begin{tabularx}{\textwidth}{llXX}
\hline
\textbf{Parameter} & \textbf{Definition / Equation} & \textbf{Value / Range} & \textbf{Distribution / Description} \\
\hline
$a$ & Dispersion constant & $4.148\,808~\mathrm{GHz^2\,cm^3\,pc^{-1}\,ms}$ & Fixed \\
$\nu_\mathrm{ref}$ & Reference frequency & 600~MHz & Fixed \\
$\nu_\mathrm{min},\,\nu_\mathrm{max}$ & Band edges & 400--800~MHz & Fixed \\
$N_\mathrm{freq}\!\times\!N_\mathrm{time}$ & Array dimensions & 512 × 7340 (simulated), resized to 512 × 1024 for model input & Fixed \\
DM & Dispersion measure & 0--600~pc\,cm$^{-3}$ & Uniform \\
$t_\mathrm{samp}$ & Sampling time & -- & Fixed (1.67~ms) \\
$\mathrm{DM}_\mathrm{max}$ & Max.\ representable DM & From Eq.~(\ref{eq:DMmax}) & Determined by array size and $t_\mathrm{samp}$ \\
$t_\mathrm{w}$ & Intrinsic pulse width & $\mu=0.0015$~s, $\sigma=0.0003$~s & Normal \\
$W$ & Effective pulse width & $\sqrt{t_\mathrm{w}^2+t_\mathrm{samp}^2+\Delta t_\mathrm{DM}^2}$ & Combines intrinsic, instrumental, and dispersion broadening \\
$t_0(\nu)$ & Arrival time & $a\,\mathrm{DM}\!\left(\frac{1}{\nu^2}-\frac{1}{\nu_\mathrm{ref}^2}\right)-t_\mathrm{w}\!\left(\frac{\nu-\nu_\mathrm{max}}{A_0\nu_\mathrm{ref}}\right)$, $A_0=0.1$ & Frequency-dependent delay and drift \\
$g(t)$ & Gaussian pulse profile & $\exp[-(t-t_0)^2/(2W^2)]$ & Defines intrinsic temporal shape \\
$\tau_\mathrm{s}$ & Scattering timescale & $\tau_0(\nu_\mathrm{ref}/\nu)^4$ & $\tau_0$ log-uniform; $\nu^{-4}$ scaling \\
$h(t)$ & Scattering kernel & $(1/\tau_\mathrm{s})\,\exp(-t/\tau_\mathrm{s})$ & Produces exponential pulse tail \\
$A(\nu)$ & Scintillation modulation & $\cos[2\pi N_\mathrm{scint}(\nu_\mathrm{ref}/\nu)^2+\phi_\mathrm{scint}]$ & $N_\mathrm{scint}$ log-uniform$[10^{-3},7]$; $\phi_\mathrm{scint}$ uniform$[0,2\pi]$; half-wave rectified \\
$P(x)$ & Gaussian noise per pixel & Mean 0, std.\ 1 & Normal \\
$F$ & Fluence scaling & -- & Inverse-uniform sampling ($F \propto u^{-2/3}$), equivalent to $p(F) \propto F^{-5/2}$\\
$\gamma$ & Spectral index & $[-4,4]$ & Uniform \\
SNR & Signal-to-noise ratio & $[\mathrm{SNR}_{\min},\mathrm{SNR}_{\max}]$ & Selected (events retained after SNR cut) \\
\hline
\end{tabularx}
\end{table*}

The parameters listed above, together with the models described in Section~\ref{sec:data-simulation}, fully determine the temporal and spectral structure of the synthetic bursts used for training and testing.


\section{Cross-validation Results}\label{app:crossvalidation}

This appendix presents the detailed results of the six-fold cross-validation used to train and evaluate the three deep-learning architectures described in Section~\ref{sec:training-performance}. 
For each model, the tables below list the fold-by-fold values of the MAE, RMSE, and MErrP for both the training and validation datasets. 
The averages reported in the main text correspond to the mean values across these folds. 
These results illustrate the consistency of convergence across folds and demonstrate the absence of significant overfitting during training.

\begin{table}
\centering
\caption{
Six-fold cross-validation performance for the three deep-learning models. 
For each architecture, the table lists the best-performing epoch per fold for both training and validation datasets. 
Metrics include the MAE, RMSE, and MErrP.}
\label{tab:AllModelsTraining_app}
\setlength{\tabcolsep}{3.5pt} 
\renewcommand{\arraystretch}{1.05}
\footnotesize
\begin{tabular*}{\columnwidth}{@{\extracolsep{\fill}}lccc@{\hspace{5pt}}ccc@{}}
\hline
 & \multicolumn{3}{c}{\textbf{Training}} & \multicolumn{3}{c}{\textbf{Validation}} \\
\cline{2-4} \cline{5-7}
Fold & MAE & RMSE & MErrP & MAE & RMSE & MErrP \\
\hline
\multicolumn{7}{l}{\textbf{Baseline CNN}} \\
\hline
1 & 0.5277 & 1.0151 & 3.24 & 0.8014 & 1.2543 & 1.47 \\
2 & 0.7505 & 1.5955 & 5.06 & 0.7430 & 1.8176 & 1.37 \\
3 & 0.5691 & 1.2270 & 5.11 & 0.6430 & 1.4148 & 1.01 \\
4 & 0.5003 & 0.9306 & 5.31 & 0.5976 & 1.5114 & 0.88 \\
5 & 0.5615 & 1.2497 & 3.84 & 0.6568 & 1.2329 & 4.96 \\
6 & 0.5430 & 1.1489 & 1.46 & 0.5922 & 1.4505 & 13.36 \\
Average & 0.5754 & 1.1945 & 4.00 & 0.6723 & 1.4469 & 3.84 \\
\hline
\multicolumn{7}{l}{\textbf{ResNet-50}} \\
\hline
1 & 0.5749 & 1.6419 & 2.16 & 0.5097 & 1.0336 & 1.09 \\
2 & 0.5181 & 1.3697 & 1.19 & 0.5376 & 0.8390 & 0.75 \\
3 & 0.6320 & 1.9343 & 1.39 & 0.3920 & 0.5467 & 0.46 \\
4 & 0.5090 & 1.3396 & 1.29 & 0.4266 & 0.6544 & 0.56 \\
5 & 0.5175 & 1.3505 & 0.87 & 0.4570 & 0.6020 & 1.08 \\
6 & 0.5140 & 1.2934 & 0.77 & 0.5583 & 0.7009 & 10.79 \\
Average & 0.5442 & 1.4882 & 1.28 & 0.4802 & 0.7294 & 2.46 \\
\hline
\multicolumn{7}{l}{\textbf{Hybrid CNN--LSTM}} \\
\hline
1 & 0.2212 & 0.2904 & 1.02 & 0.2674 & 0.6396 & 0.34 \\
2 & 0.2527 & 0.3581 & 1.21 & 0.2553 & 0.5941 & 0.43 \\
3 & 0.2158 & 0.2888 & 0.76 & 0.2502 & 0.7089 & 0.34 \\
4 & 0.2038 & 0.2728 & 0.84 & 0.2479 & 0.6067 & 0.32 \\
5 & 0.2117 & 0.2862 & 0.78 & 0.2415 & 0.5097 & 1.30 \\
6 & 0.1922 & 0.2527 & 0.42 & 0.2554 & 0.5831 & 2.46 \\
Average & 0.2162 & 0.2915 & 0.84 & 0.2530 & 0.6070 & 0.86 \\
\hline
\end{tabular*}
\end{table}


\section{Test-set Results}\label{app:testresults}

This appendix presents the six-fold test-set results for the three deep-learning architectures described in Section~\ref{sec:test-performance}. Each model was evaluated using MAE, RMSE, MErrP, and two accuracy indicators, the GER and GDR. The averages reported in the main text are the means across these folds.

\begin{table*}
\centering
\caption{Six-fold test-set performance of the three deep-learning architectures. Each block lists the MAE, RMSE, MErrP, as well as the GER and GDR. The final row reports the fold-averaged performance for each model.}
\label{tab:testresults_app_combined}
\setlength{\tabcolsep}{3pt}
\renewcommand{\arraystretch}{1.05}
\footnotesize
\begin{tabular*}{\textwidth}{@{\extracolsep{\fill}}lccccc@{\hspace{6pt}}ccccc@{\hspace{6pt}}ccccc@{}}
\hline
 & \multicolumn{5}{c}{\textbf{Baseline CNN}} & \multicolumn{5}{c}{\textbf{ResNet-50}} & \multicolumn{5}{c}{\textbf{Hybrid CNN--LSTM}} \\
\cline{2-6} \cline{7-11} \cline{12-16}
Fold & MAE & RMSE & MErrP & GER ($\%$) & GDR ($\%$) & MAE & RMSE & MErrP & GER ($\%$) & GDR ($\%$) & MAE & RMSE & MErrP & GER  ($\%$) & GDR ($\%$) \\
\hline
1 & 0.8019 & 1.1070 & 0.97 & 88.81 & 76.58 & 0.5093 & 0.6679 & 0.66 & 90.20 & 92.25 & 0.2595 & 0.5902 & 0.26 & 96.85 & 98.00 \\
2 & 0.7461 & 1.4244 & 0.94 & 90.27 & 81.56 & 0.5391 & 0.6775 & 0.53 & 93.48 & 91.58 & 0.2478 & 0.5218 & 0.27 & 96.64 & 98.42 \\
3 & 0.6363 & 1.1047 & 0.71 & 93.24 & 85.51 & 0.3949 & 0.5426 & 0.39 & 94.78 & 95.52 & 0.2667 & 1.0382 & 0.28 & 96.46 & 97.77 \\
4 & 0.5915 & 0.9358 & 0.70 & 93.77 & 86.03 & 0.4219 & 0.5510 & 0.43 & 92.79 & 95.27 & 0.2467 & 0.5258 & 0.24 & 96.83 & 98.02 \\
5 & 0.6581 & 1.0920 & 0.68 & 93.74 & 83.22 & 0.4621 & 0.6171 & 0.41 & 93.54 & 93.34 & 0.2441 & 0.5972 & 0.26 & 96.87 & 98.19 \\
6 & 0.5799 & 0.9633 & 0.65 & 93.52 & 87.60 & 0.5542 & 0.6955 & 0.83 & 89.23 & 90.85 & 0.2605 & 0.5842 & 0.26 & 96.40 & 97.94 \\
Average & 0.6690 & 1.1045 & 0.77 & 92.23 & 83.42 & 0.4802 & 0.6253 & 0.54 & 92.34 & 93.13 & 0.2542 & 0.6429 & 0.26 & 96.68 & 98.06 \\
\hline
\end{tabular*}
\end{table*}

\bsp	
\label{lastpage}
\end{document}